\documentstyle[twocolumn]{mn}
\def\beq{\begin{equation}}
\def\eeq{\end{equation}}
\def\bey{\begin{eqnarray}}
\def\eey{\end{eqnarray}}

\def\Max{\rm{Max}}
\def\Min{\rm{Min}}
\input epsf

\title[]
	{Origin of non-unique deprojection of the Galactic bar}
\author[]
	{HongSheng Zhao
	\thanks{Sterrewacht Leiden (hsz@strw.LeidenUniv.nl)}}
\pagerange{\pageref{firstpage}--\pageref{lastpage}}
\pubyear{1997}

\begin{document}
\maketitle
\label{firstpage}

\begin{abstract}

Non-uniqueness of deprojecting the integrated light distribution of a
nearby or faraway triaxial body is reviewed in the context of deriving
the volume density of the Galactic bar from the COBE/DIRBE maps of the
Galactic plane.  The exact origin of this non-uniqueness is studied.
One can write down a sequence of triaxial bar models which appear
identical in integrated light from the Sun's perspective, and the
whole sequence is mapped out as a function of the Galactocentric
distance of the observer from galactic to extragalactic distance
scales.  While mirror symmetries and perspective effects are
compatible with any orientation of the bar in the sequence, weak upper
and lower bounds can still be placed on the angles and axis ratios of
the bar by positivity and other general requirements.  Star count data
of bulge giants are ideal for selecting a unique model from the
sequence.

\end{abstract}


\section{Introduction}

Deprojection of galaxies from the observed light distribution on the
sky plane to the intrinsic 3D volume luminosity distribution is one of
the basic problems of astronomy.  It is common knowledge that the
deprojected results are generally non-unique because of the freedom of
distributing stars along any line of sight.  General constraints such
as symmetry, positivity and self-consistency are often too weak to
completely break the degeneracy of models, in particular, the
fundamental degeneracy of a triaxial ellipsoidal distribution vs an
axisymmetric spheroidal one; both can project to the same elliptical
isophotes in the line-of-sight integrated light distribution
(Contopoulos 1956).  Kinematic information such as rotation around the
apparent minor/major axes of the isophotes and the 2D projected
velocity field is necessary to solve this so-called ``shape problem''
of elliptical galaxies (Binney 1985, Franx, Illingworth \& de Zeeuw
1991, Statler 1994 and references therein).

Nevertheless, as Blitz \& Spergel (1991) pointed out in the case of
the Galactic bar vs. an oblate bulge, the degeneracy is at least
partially broken by photometric information from the perspective
effect of nearby triaxial or bi-symmetric objects, namely with mirror
symmetry with respect to three orthogonal planes.  Perspective is the
basis for the detection of the Galactic bar in the maps of the
Galactic plane made by the DIRBE experiment aboard the COBE satellite.
The dust-corrected infrared maps from COBE show significant systematic
brightness variations between the right hand side ($l>0^o$) and the
left hand side ($l<0^o$), and small systematic differences between the
lower ($b<0^o$) and the upper ($b>0^o$) Galactic plane (Weiland et
al. 1994): a strong signature of a nearly edge-on (mirror symmetry
with the $b=0^o$ plane) triaxial bar with the major axis of the bar
rotated by an angle from our line of sight.  A simple sketch of the
geometry is shown in the top left diagram of Fig.1).

Dwek et al. (1995) made the first attempt to fit the COBE/DIRBE maps
with a few simple analytical luminosity models.  They found that the
COBE maps are grossly consistent with several triaxial models: a very
strong bar pointing almost along the Sun-Galactic center line of sight
or a short bar pointing more sideways.  Despite still significant
residues of all their models, the existence of reasonable fits of the
COBE map with a wide range of bar angles, axis ratios, radial profiles
and degrees of boxyness suggest an intrinsic non-uniqueness among
triaxial models which is not lifted by the perspective effect.

Binney \& Gerhard (1996) separated the non-unique orientation of the
bar from the likely unique 3D distribution once the three mirror
planes of the bar are fixed.  Their numerical experiments show that
the final result of their non-parametric deprojection depends only on
the assumed orientation of the bar, but not on the initial seed models
for starting their Lucy-Richardson iterations.  They explain the
latter uniqueness as a compromise between the symmetry requirements of
the density and a good fit to the perspective-induced asymmetries in
the COBE map.

Binney, Gerhard \& Spergel (1997) applied this non-parametric method
to the COBE map after correcting for dust with an advanced extinction
model from Spergel, Malhotra \& Blitz (1997).  They concluded that
the major axis angle of the bar is likely very close to $20^o$ with a
good fit to the COBE map still possible for a  positive
smooth triaxial bar with major axis angle between $15^o$ and $35^o$.  They
suspect that these limits on the bar angle mark the boundary of
unphysical densities with negative regions and/or peculiar profiles.

Compared to external systems, perspective provides the unique
signature for the Galactic bar as well as the subtleness to its
non-uniqueness.  The non-uniqueness in a nearby bar is entangled with
perspective effects, making it a to-be-defined narrow sub-class of
non-uniqueness for external systems.  Simple transformations such as
shearing/stretching an external ellipsoid in the line-of-sight
direction can generate a sequence of ellipsoids with the same surface
brightness map (Stark 1977, Franx 1988).  But as Binney \& Gerhard
(1996) emphasize, moving material along the line of sight is forbidden
by the mirror symmetries of a nearby bar.  The well-known trick with
axisymmetrical galaxies of adding a small amount of unphysical density
with zero surface brightness (Rybicki 1986, Palmer 1994, Kochanek \&
Rybicki 1996, van den Bosch 1997, Romanowsky \& Kochanek 1997),
christened konuses by Gerhard \& Binney (1996) for extragalactic
axisymmetric bodies would not apply either.  While preserving the
orientation and the symmetry of the bar, adding konuses perturbs the
left-to-right difference map.

Unfortunately none of the previous analyses with the COBE bar and
external bars offer a satisfactory explanation for the non-uniqueness.
In particular, the true source of non-uniqueness is never clearly
singled out, and how this non-uniqueness varies as a function of the
distance to the body is never clearly stated.  It is also unclear how
to transform one solution to another and map out the sequence of
models admitted by the COBE maps, which is one of the main sources of
information on the stellar distribution of the Galactic bar.  Answers
to such questions are important for integrating the constraints from
the COBE map to other data of the bar from star counts, gas/stellar
kinematics, and microlensing (see de Zeeuw 1993, Gerhard 1995).  For
example, the angle of the bar is a main source of uncertainty in
comparing the COBE map with observed microlensing optical depth of the
bar (Zhao \& Mao 1996, Bissantz, Englmaier, Binney \& Gerhard 1997);
it is also a main source of uncertainty in determining the number
density of brown dwarfs in the Galactic bar using microlensing (Zhao,
Rich and Spergel 1996).

This paper gives an analytical description of the nature of the
non-uniqueness in the nearby COBE bar and points out its main
differences with non-uniqueness in external systems.  I also show how
the positivity of the distribution puts a loose limit on the bar
models, and discuss the best methods to lift the remaining degeneracy.

\section{The nature of non-uniqueness}

In this section I will describe the sources of non-uniqueness purely
in words and illustrations without getting into the more mathematical
aspect of the problem, which will be supplimented in \S3 together with
some applications to the COBE bar.

\subsection{Mirror symmetries vs. invisible densities}

Consider for the time being only bars which are symmetric with respect
to the $b=0^o$ plane.  Divide the Galactic bar and the COBE map into a
left and a right part with the $l=0^o$ plane, which passes the
Sun-center line and the rotation axis of the Galaxy.  When folded
along the $l=0^o$ line, the COBE/DIRBE Galactic plane map $I(l,b)$ can
be decomposed into two independent maps: a difference map
$I(l,b)-I(-l,b)$ by subtracting the $l<0^o$ side from the $l>0^o$
side, and an addition map $I(l,b)+I(-l,b)$ by adding up the two sides.
These two maps are line-of-sight integrations of the odd and the even
parts of the volume density $\nu$ of the inner Galaxy, $\nu_{odd}$ and
$\nu_{even}$, decomposed in terms of the symmetry with respect to the
$l=0^o$ plane.  The difference map and $\nu_{odd}$ are anti-symmetric
with $l=0^o$ plane, the addition map and $\nu_{even}$ are
left-to-right symmetric.

{\sl Non-uniqueness originates from the large degrees of freedom in
fitting the left-to-right symmetric addition map $I(l,b)+I(-l,b)$}
\footnote{Deprojection of the difference map $I(l,b)-I(-l,b)$, which
distinguishs a bar from an axisymmetric bulge, is {\sl assumed} to
give a unique odd solution $\nu_{odd}$ in this paper.}: any even
component, such as an axisymmetric bulge or disc, a long end-on bar
(the major axis coincides with the Sun-center line), a short side-on
bar (the middle axis coincides with the Sun-center line) look very
much alike in projection.  In fact, in subtracting the axisymmetric
component from the end-on bar model, one makes an unphysical
distribution with some negative density regions, which is similar to
extragalactic konuses in the sense that its projected light intensity
is exactly zero after integration over any line of sight.  I shall
call such left-to-right symmetric unphysical components ``invisible
densities'' since superimposing them on the bar is undetectable in
projected light.  They are the sources of non-uniqueness.

\begin{figure}
\epsfysize=3.5cm
\leftline{\epsfbox{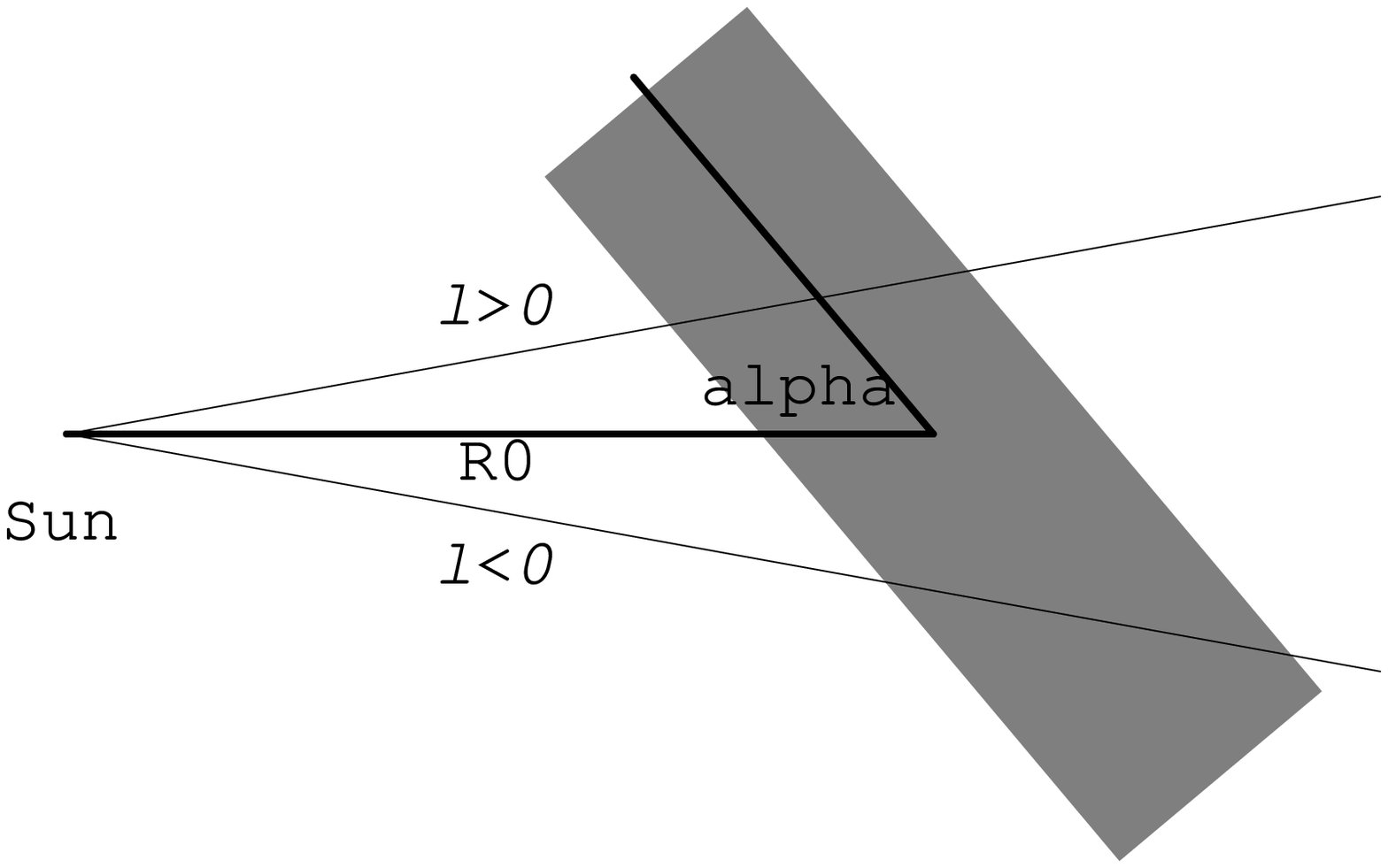}}
\vskip -3.5cm
\epsfysize=3.5cm
\rightline{\epsfbox{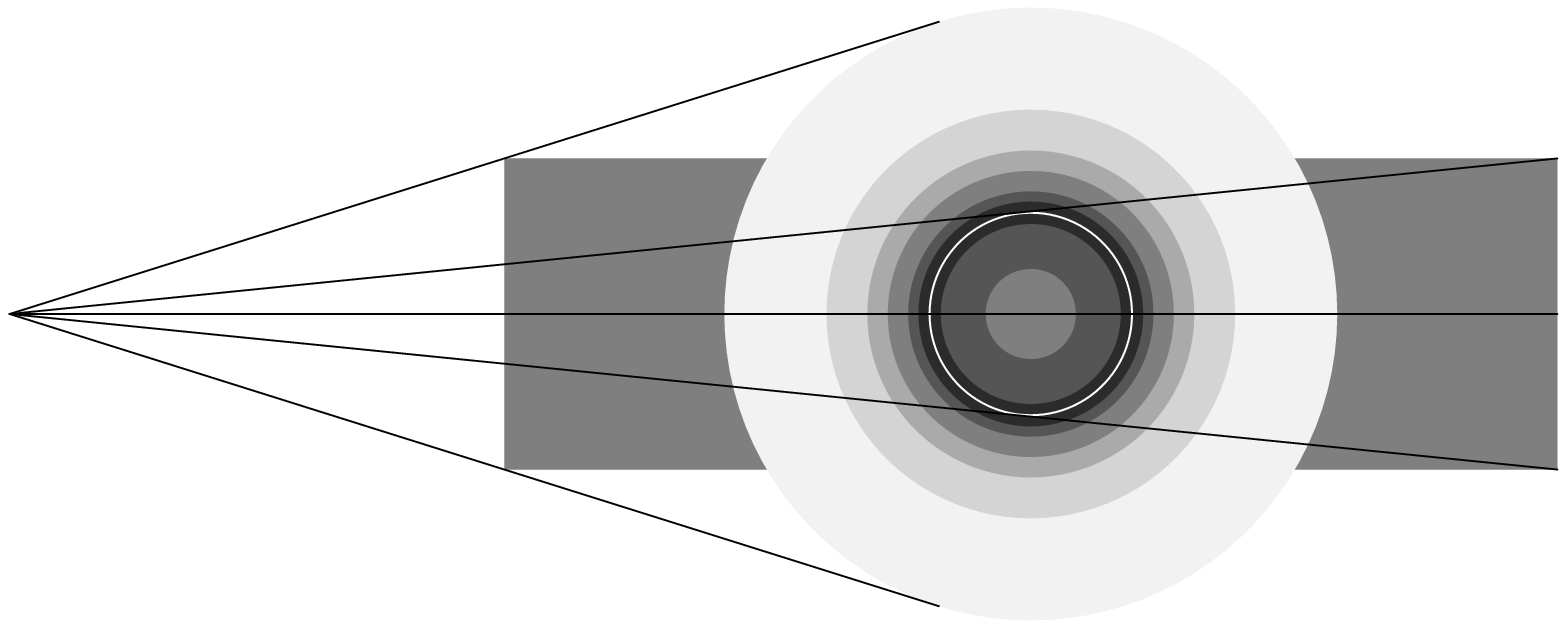}}
\epsfysize=7cm
\epsfbox{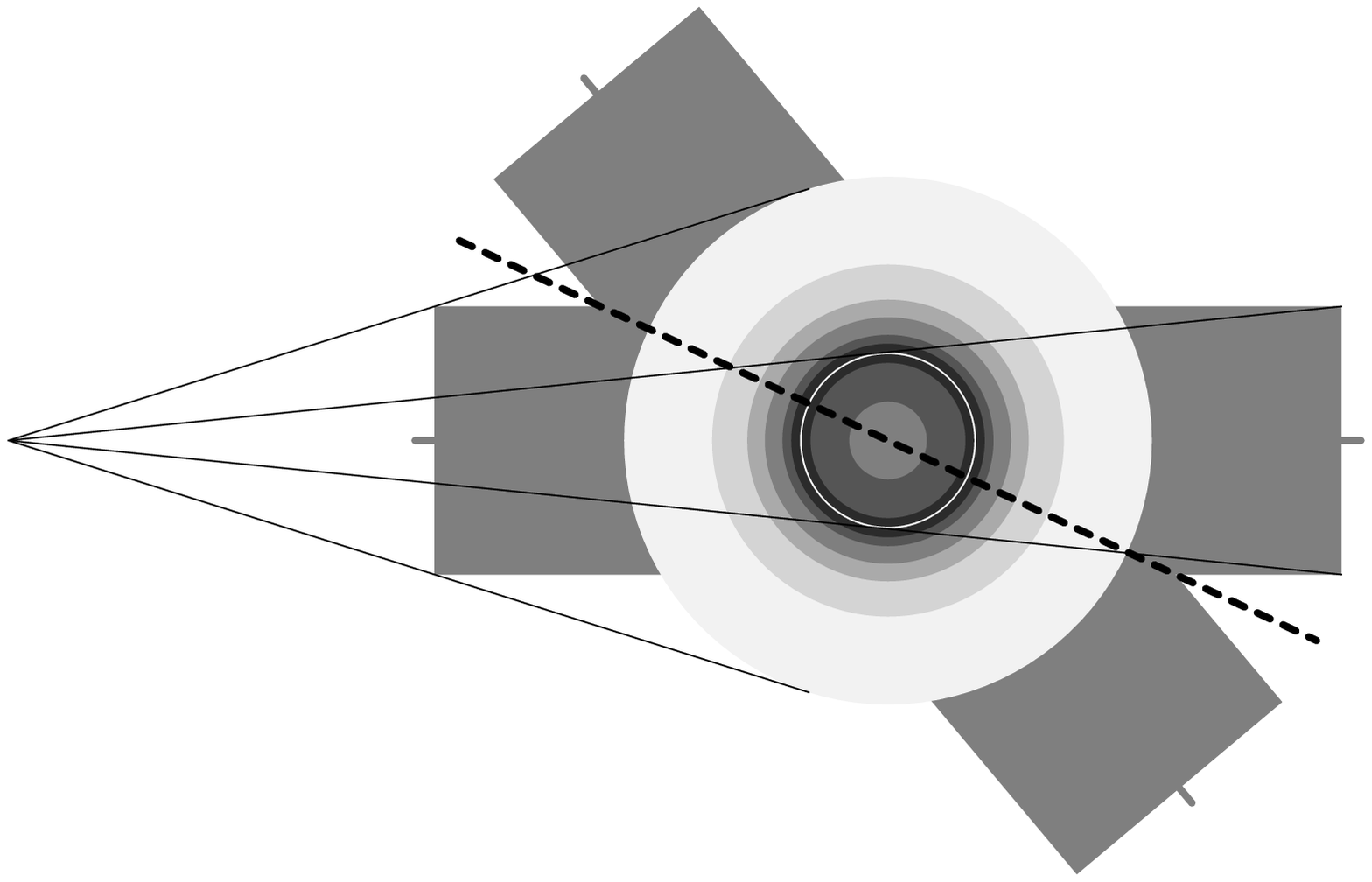}
\caption{
Crossections of a uniform cigar-shaped bar tilted at an angle $\alpha$ from
the Sun-center line (upper left), a two-component invisible density
(upper right), and a superposition of the above two bodies 
(the enlarged lower diagram); several line-of-sight pathes from the Sun
are drawn in thin solid lines.  The invisible density is made by
subtracting from an end-on cigar-shaped bar a stratified ball-shaped bulge of
the same angular size and projected intensity.  
Adding such an invisible density to the cigar-shaped bar on the upper left
has no effect on the latter's projected intensity, but rotates
the mirror plane of the spatial distribution 
to the dotted line in the middle at an angle ${\alpha \over 2}$
from the Sun-center line.}
\end{figure}

An example of an invisible density is illustrated in the upper right
corner of Fig.1, where I tailor the radial profile of a stratified
ball-shaped bulge in such a way so to have the same angular size and
projected intensity as a uniform cigar-shaped bar placed end-on.
Subtracting the ball-shaped bulge from the cigar-shaped bar yields an
invisible density.  The thin dark ring in the bulge is a density peak,
corresponding to the line of sight to the far edge of the cigar-shaped
bar, a direction where the depth and the projected intensity of the
bar are at maximum.  The fall-off of density towards the edge of the
bulge corresponds to the ever-decreasing depth of the cigar-shaped bar
with increasing impact parameter of the line of sight.  Now
superimpose a finite fraction, say $f$, of this unphysical component
on the COBE bar, say, it is an ellipsoidal bar with a Gaussian radial
profile.  While such an ``operation'' is invisible in the projected
light, its result is a new density with generally so irregular
isophotes that there is no specific symmetry plane.

When the mirror symmetries with respect to three given orthogonal
planes are imposed on the density distribution of the bar, the
forementioned ``invisible operation'' is nearly forbidden because of
the general twist of the major axis position, but not always.  The
exception is, as shown in the enlarged lower diagram of Fig.1, when
the superimposed end-on bar is a clone of the original bar, in which
case the final density should have mirror symmetry with respect to a
new plane (the dotted line) which divides the angle of the Sun-center
line and the major axis of the original bar in half.  Namely the
mirror symmetry is preserved, only the symmetry plane is different.
Subtracting the round bulge has no further effect on the symmetry
plane, but will take away any trace of transformation in the addition
map of the integrated light.  So {\sl an invisible density which
rotates and preserves mirror symmetry} is obtained by placing the
original bar at end-on then subtracting off a round bulge.  Generally
the new major axis position depends on the fraction $f$ of the added
invisible density.  When $f$ is sufficiently small, the new density
could also be positive everywhere.  The general technique of designing
invisible densities for a realistic triaxial bar is shown in \S3.1 and
\S3.3, in particular, how to rotate the bar without twisting its major
axis with radius.

\subsection{Perspective effect, extinction and best-fit models}

How is the perspective effect of the bar decoupled from the
non-uniqueness problem?  The perspective effect is only prominent in
the difference map, and is by-passed by restricting all the
transformations to the even part of the density: $\nu_{odd}$ is never
affected by the invisible density.  Still the invisible density is a
function of the Galactocentric distance of the observer $R_0$.  For a
hypothetical observer receding from the Galactic bar towards the
anti-center, its Galactocentric distance $R_0$ increases, (the
axisymmetric part of) the invisible density should be adjusted
slightly to accomodate the changing perspective so to preserve this
kind of non-uniqueness all the way to extragalactic distance.  The
common ambiguity of an extragalactic axisymmetric bulge with an end-on
bar is a very special case of the kind of non-uniqueness with the COBE
bar.  As I will come back to in \S3.2, {\sl other distinctive features
of the invisible density are that it has a net luminosity, it is not
confined to any cone in the Fourier $k$-space and it is symmetric with
respect to the $l=0^o$ plane}.

Will invisible density remain invisible in the dusty Galaxy?  Even at
the near infrared the extinction of the COBE bar is a strong source of
uncertainty for deprojection as shown by the strong color variation in
the COBE/DIRBE maps (Arendt et al. 1994).  In the first order
approximation where dust is primarily situated in the foreground within
the near 2 kpc of the line of sight to the center (Arp 1965), the
projected intensity of the bar is simply multiplied by a reddening
factor, thus the zero projected intensity of the invisible density is
not affected.  However at low latitude ($|b| \le 3^o$) where dust and
stars are mixxed both in the disc and in the bar regions (Spergel,
Maholtra \& Blitz 1997), extinction will
modulate the bar density with an ill-determined function for the dust
distribution, which should be deprojected simultaneously with the
stellar distribution.  The invisible density, defined in the dust-free
context, will cast faint (irregular) shadows on the COBE/DIRBE maps;
one sees the near end of the end-on cigar-shaped bar (cf. upper right
of Fig. 1) faintly in a heavy mist of interstellar dust.  Although in
principle one could still define a certain density which projects to
zero after being modulated by dust, and infer the dust distribution
from, say, star count data, independent of the to-be-deprojected maps,
but finding such a density which preserves the triaxial symmetry of
the bar becomes hopelessly difficult.

Do ``best-fit'' bar models mean anything?  Since for any bar which
fits the COBE/DIRBE maps, one can build a sequence of triaxial 
models with any bar angle to fit the data with exactly the same
accuracy by adding different amount of the invisible density, {\sl it
is ill-defined to speak of the ``best-fit'' bar model or bar angle
from the COBE/DIRBE maps}.  The ``best-fit'' bar angles derived in the
parametrized models of Dwek et al.  are model dependent
determinations.  These, which could have fairly small error bars
and/or residues (e.g. their G1, G2 and E3 models), are a mere
reflection of the different ranges where their different {\it a
priori} assumptions of the boxiness and radial profile of the density
are ``best'' satisfied.  The systematic residues in their models
reflect only the unphysicalness of the {\it a priori} assumptions in
detail (their diamond shaped bar is the best example).  In fact it
would not even be surprising to have a perfect fit to the data for
arbitrary bar angle (cf. eqs.~\ref{eveni2}-~\ref{oddi2}) provided the
dereddened COBE maps are sufficiently smooth because there are in
principle enough degrees of freedom in the mathematical functional
forms of a triaxial density to make the problem underconstrained.  It
is straightforward to show that for observers nearby or at infinity
this kind of ambiguity with the major axis angle occurs in every even
$m$ mode ($m=2,4,...$), odd $m$ mode ($m=1,3,...$), and m-fold spiral
arm mode where the major axis angle $\alpha$ also changes with $R$,
and also in any band-limited superposition of finite number of modes.
In all cases the major axis angle cannot be determined solely on the
basis of the goodness of fit to the COBE map.

Then where do the limits for the bar angle come from?  The only hope
left to constrain the bar angle from the COBE maps is to impose some
generally sensible unbiased requirements of the density, such as
positivity, and to some extent smoothness, regular shape and radial
profiles, as in Binney, Gerhard \& Spergel (1997).  An example is
given in the following to illustrate the main points in this section.

\section{Application to the COBE bar}

So far I have assumed that one can design invisible densities which
project strictly to zero even for nearby bodies, and I have also
restricted to the simple case that the observer is in the mid-plane of
the bar.  In this section some examples of the invisible density are
written down as explicit functions of the observer's distance and
viewing angles to the bar, allowing for a small vertical offset of the
Sun from the mid-plane of the bar (Dwek et al. 1995, Binney, Gerhard
\& Spergel 1997).  These toy models are used both to illustrate the
non-uniqueness of the COBE bar and and to study possible methods to
break the degeneracy.

\subsection{An $m=2$ bar viewed from a general distance and perspective}

An $m=2$ triaxial density distribution with any
 orientation of symmetry planes can be 
fairly generally written in a Galactocentric polar coordinate
$(r,\theta,\phi)$ as
\bey\label{nuf0} 
\nu_{f=0}(r,\theta,\phi)
&=& G(r) + P(r) H_{f=0}(\theta,\phi),
\eey
which is a superposition of 
a spherical component  $G(r)$, and a triaxial perturbation
with the radial profile $P(r)$ and the angular part 
$H_{f=0}(\theta,\phi)$; the latter equals to
a function $H_f(\theta,\phi)$ evaluated at $f=0$, where
\bey\label{Hf}
H_f(\theta,\phi) &\equiv &  s_0 + (f+c_0) \cos^2 \theta  \\\nonumber
& & + 2 (c_{1} \cos \phi + s_{1} \sin \phi) \sin \theta \cos \theta \\\nonumber
& & + (c_{2} \cos 2\phi + s_{2} \sin 2\phi) \sin^2 \theta.
\eey
One can show that the perturbation is triaxial since  
$r^2H_{f=0}(\theta,\phi)$ reduces to a quadratic function in 
rectangular coordinates with ellipsoidal isosurfaces whose 
symmetry planes are determined by the constants
$c_0$, $c_1$, $c_2$, $s_0$, $s_1$, and $s_2$.

Now a sequence of triaxial models with the same projected density as
$\nu_{f=0}(r,\theta,\phi)$ can be made if one can write down its
invisible density, which should be made by subtracting a spherical
model from an end-on prolate bar with amplitude $P(r)$.
For convenience fix the pole of the Galactocentric polar coordinates 
$(r,\theta,\phi)$ as the Sun-center line of sight, and 
the zero point of the roll angle 
$\phi$ as the $b=0^o$ plane, so that in this coordinate system
$P(r) \cos^2 \theta$ would describe an end-on prolate 
(i.e., $\phi$-independent) distribution.  
Now suppose a spherical density $S(r)$ can be inverted$^{\ddagger}$
from the integral equation,
\beq\label{tiltk}
\int_0^\infty \!\! S(r) dD = \int_0^\infty \!\left[P(r)\cos^2\theta\right]dD \equiv T(l,b),
\eeq
where $\int_0^\infty \!\! dD$ is an integration 
along any line-of-sight direction $(l,b)$,
then
\beq
\nu_f(r,\theta,\phi) =\nu_{f=0}(r,\theta,\phi) +
\left[P(r) \cos^2 \theta -S(r) \right] f,
\eeq
describes a sequence (as function of the parameter $f$) of triaxial models 
with the same projected density.  The sequence clearly preserves
triaxial symmetry because like $\nu_{f=0}(r,\theta,\phi)$ of eq.~\ref{nuf0}
\bey\label{nuf1} 
\nu_f(r,\theta,\phi)
&=& \left[G(r) - f S(r) \right] + P(r) H_f(\theta,\phi),
\eey
also consists of a spherical component $G(r) - f S(r)$, and
a triaxial perturbation
with an angular dependence $H_f(\theta,\phi)$ similar to
$H_{f=0}(\theta,\phi)$ (cf. eq.~\ref{Hf}) except that
the orientation of the symmetry planes will also depend on $f$.

By tuning solely $f$ the model (cf. eq.~\ref{nuf1}) changes
perspective with the Sun: the offset of the Sun from the mid-plane and
the rotation of the major axis in the bar mid-plane are varied
together along a 1D sequence.  The less interesting third angle which
describes a roll of the bar mid-plane around the Sun-center line, and
is equivalent of the sky position angle in the case of extragalactic
bodies, is not affected by adding an invisible density; the latter is
invariant of a roll around the line-of-sight axis.

As a specific example to show how the model changes with
the observer's distance and perspective, let
\beq\label{pr}
G(r)=\nu_0 \exp(-{r^2 \over 2 a^2}),~~~
P(r)= \nu_0 {r^2 \over a^2} \exp(-{r^2 \over a^2}),
\eeq
where $a$ is the characteristic scale of the model.
The spherical density 
$S(r)$, as specified in the integral eq.~\ref{tiltk}, can be solved
analytically with a variation of the well-known Eddington formula 
for deprojecting spherical system.
\footnote{
Generally if $J(p)$ is the integrated intensity of the light distribution 
$S(r)$ or $P(r) \cos^2\theta$
along a path with a Galactocentric impact parameter $p$, including
the contributions from both the forward direction and 
the backward direction, $T(l,b)$ and $T(\pi+l,-b)$,
then 
$J(p) \equiv T(l,b)+T(\pi+l,-b) \equiv
\int_{-\infty}^{\infty} \! \! S(r) du 
= \int_{-\infty}^{\infty} \! \! P(r) \cos^2 \theta du$, where
$u$ is the offset along the line of sight from the tangent point, 
$r=\sqrt{p^2+u^2}$, and 
${ r \cos\theta \over p} = \left( {u \over p} \sqrt{1-{p^2 \over R_0^2}} 
              - {p \over R_0} \right)$.
$S(r)$ is then inverted with an Abel transformation
$S(r)=-\pi^{-1} \int_{r}^{\infty} \! \! {dJ(p) \over dp} (p^2-r^2)^{-1/2} dp$.}
The dependence on the Galactocentric distance $R_0$ is generally such that
if $S_\infty(r)$ is $S(r)$ in the limit of $R_0 \rightarrow \infty$
then $S(r)-S_\infty(r) \propto \left({a \over R_0}\right)^2$.
For the models in eq.~\ref{pr},
\beq\label{sr}
S(r)= S_{\infty}(r) \left[ 1 + {a^2 \over R_0^2}
\left({2r^4 \over a^4} - {3 r^2\over a^2} \right)\right],~~
S_{\infty}(r) = {\nu_0 \over 2} \exp(-{r^2 \over a^2}),
\eeq
and
\bey\label{nuf}
\nu_f(r,\theta,\phi) &=& \nu_0 \exp(-{r^2 \over 2 a^2}) \{
1 - \exp(-{r^2 \over 2 a^2}) 
 [ \\\nonumber
& & {f \over 2} + f{r^2 \over R_0^2}
\left({r^2 \over a^2} - {3 \over 2} \right)
- {r^2 \over a^2}H_f(\theta,\phi)
] \},
\eey
where the extra damping factor 
$\exp(-{r^2 \over 2 a^2})$ in front of the square bracket is to
guarantee that the model is spherical and positive at large radius.
The above equation yields triaxial
models with monotonic and positive density
and various degrees of flattening/triaxiality and boxyness
for a wide range of scale length $a$ and shape and perspective parameters
$(c_0,c_1,c_2,s_0,s_1,s_2)$ and $f$.  The models resemble  
the $G1$ and $G2$ models of Dwek et al. (1995) for the Galactic bar 
in terms of a Gaussian radial profile.  
Since the Sun is only about $10$ pc from the midplane of the bar
(Binney, Gerhard \& Spergel 1997), the parameters $s_1$ and 
$s_2$ should be of the order $10{\rm pc}/R_0 \sim 10/8000 \sim 0.01$,
much smaller than unity.
In particular, models with
\beq\label{edgeonf}
s_1=s_2=0,~~f=c_2-c_0 - 2 c_1 \cot 2 \alpha,
\eeq
prescribe a sequence of edge-on (mirror symmetry with $b=0^o$ plane) 
$m=2$ perturbation, where $\alpha$ is the angle
between the major axis of the perturbation and
the $l=0^o$ plane which passes the Sun and the rotation axis of the Galaxy.
Fig.2 shows an equatorial slice of two such edge-on models
from the sequence at $\alpha=25^o$ and $50^o$ respectively.
The shape of bar is a function of the major axis angle such that 
models in a sequence yield
the same projected light intensity from the Sun's perspective.

\begin{figure}
\epsfysize=7cm
\epsfbox{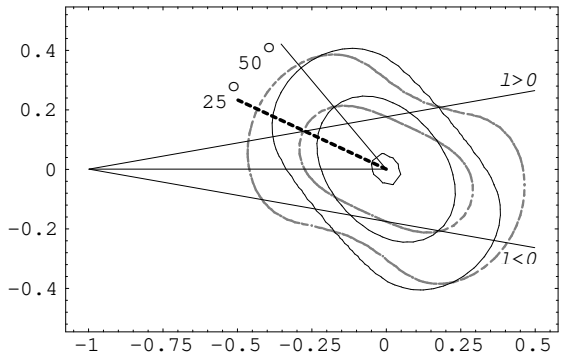}
\caption{
Equatorial slices of two triaxial 
models with identical surface brightness map, one
at $\alpha=25^o$ (thick dashed contours) and the other at $\alpha=50^o$ 
(thin solid contours).  The Sun is at $(-1,0)$ to the central left. 
Two solid lines are drawn to show a line of sight at positive and
negative longitude respectively.}
\end{figure}

\subsection{Non-uniqueness of the COBE bar vs. external systems}

Now I digress from the COBE models to stress a few key differences
between the non-uniqueness of the COBE models and external systems.
{\sl Unlike its external counterpart with $R_0 \rightarrow \infty$,
the invisible density for a nearby triaxial body is not massless.}
For the bar models here, the invisible density has a mass 
proportional to ${a^2 \over R_0^2}$.
The total luminosity of the bar models (cf. eqs.~\ref{nuf} and~\ref{Hf})
\beq\label{lum}
L \equiv \int\!\!\! d^3r \nu(r,\theta,\phi) = \pi^{3 \over 2} \nu_0 a^3 \left[2^{3 \over 2} + {3 s_0 + c_0 \over 2} -{3 \over 2} {a^2 \over R_0^2} f \right],
\eeq
changes with $f$.

To fit the same projected light intensity the models should have the
same normalization $\nu_0$, but the total luminosity changes by a
factor on the order of ${a^2 \over R_0^2} \sim$ a few percent for the
COBE bars.  This is because unlike external systems, the observed
angular distribution of light in the COBE bar does not simply sum up
to a unique measurement of its total luminosity.  The result depends
on knowing the distance to the bar and the orientation of the bar.
The upper panel of Fig.3 shows that an observer at the Sun's position
$R_0=8$kpc may slightly underestimate/overestimate the intrinsic
luminosity by about $10\%$ depending on assumptions of the bar angle.
Moving out to the anti-center the observer 
has a full outside view of the bar 
and hence a better determination of its luminosity
but sacrifices somewhat tighter constraint on the bar angle from 
the perspective effect and positivity.  Going to the center of the
bar, the observer has full information of the orientation of the bar
(the brightest direction is the major axis direction), but very poor
information of the intrinsic luminosity profile and total luminosity.

Another feature of {\sl the invisible density here  is that unlike
konus-like structures, it is not confined to any cone in the Fourier
{\bf k}-space even in the limit $R_0 \rightarrow \infty$.}
This is shown explicitly by
computing the Fourier transform of the (prolate band-limited) 
invisible density (cf. eqs.~\ref{pr} and ~\ref{sr}) using
eqs. 4, 24 and 25 of Palmer (1994).  
\bey\label{kspace}
F(\bf{k}) &\equiv &
\int \! \! d^3r \exp(-i {\bf k} \cdot {\bf r} )\left[P(r) \cos^2\theta -S(r)\right]\\
&=&  F_{R_0 \rightarrow \infty} ({\bf k}) + {a^2 \over R_0^2} T(k),
\eey
which is a {\bf k}-space prolate distribution around the Sun-center line,
where
\beq\label{kinfty}
F_{R_0 \rightarrow \infty} ({\bf k})
\equiv \pi^{3 \over 2}  \nu_0 a^3 
\left(\cos^2 \theta_{{\bf k}} -{2 \over 3}\right)
{k^2a^2 \over 4}\exp(-{k^2a^2 \over 4}),
\eeq
$\theta_{{\bf k}}$ 
is the angle of the {\bf k}-vector with the Sun-center line,
and
\beq
T({\bf k})  \equiv  \pi^{3 \over 2}  \nu_0 a^3 
\left(-{3 \over 2} + {21 k^2a^2 \over 24} -{k^4a^4 \over 16} \right) 
\exp(-{k^2a^2 \over 4}).
\eeq
One recovers the mass of the invisible 
density in the limit ${\bf k} \rightarrow 0$,
$F(0)={a^2 \over R_0^2}T(0)=-{3 a^2 \over 2 R_0^2} (\pi^{3 \over 2}\nu_0 a^3)$ 
(cf. eq~\ref{lum}).
Eq.~\ref{kinfty} shows that even when the bar is placed at infinity,
$F({\bf k})=F_{R_0 \rightarrow \infty}({\bf k}) \ne 0$ everywhere (except 
at {\bf k}$=0$ and/or $\theta_{{\bf k}}=\cos^{-1} \sqrt{2 \over 3}$).  
Since $F({\bf k})=0$ for konus-like structures (or its triaxial version) 
outside a certain ``cone of ignorance'' around a principal axis
of the model (Rybicki 1986, Gerhard \& Binney 1996,
Kochanek \& Rybicki 1996), {\sl the kind of non-uniqueness shown in this paper
has little to do with konuses}.  The two sequences of models meet
only for an end-on prolate bar seen at infinity, which can alternatively be 
interpreted as an axisymmetric system with inclination $i=0$ (face-on).
In this special case the invisible density here becomes a konus but
the half opening 
angle of the ``cone of ignorance'' is as large as ${\pi \over 2}$.

{\sl The invisible density is also of different nature from the kind of
non-uniqueness associated with a simple shear and/or stretch
transformation of an ellipsoid}.  This applies even in the limit that
the object is at infinity because these transformations normally
change the odd part of the density, while the invisible density here
is strictly an even function.

\subsection{Limits on the bar angle vs. positivity}

The limits for the bar angles are given by the positivity requirement.
They can be derived analytically in the context
of models more general than eq.~\ref{nuf} without additional effort.
As it turns out, the results are simplest for the broad class of 
interesting edge-on models with an axisymmetric component and 
an $m$-mode perturbation; $m=2,4,...$ if 
mirror symmetry is imposed.  These models have a density distribution
which can be written with no loss of generality as
\beq\label{num}
\nu_{\alpha}(R,z,\psi) = \left[A_{\alpha}(R,z) + 
{ B_m(R,z) \over \sin m\alpha}  \cos m (\psi-\alpha)\right],
\eeq
\bey
A_{\alpha}(R,z) &\equiv& C(R,z) -\cot m\alpha V_m(R,z),\\\label{km}
K_m(R,z,\psi) &\equiv & \left[ B_m(R,z) \cos m\psi - V_m(R,z) \right],
\eey
and
\beq\label{kmin}
\int_0^\infty\! \! \!dD  K_m(R,z,\psi) =0,
\eeq
where a Galactocentric cylindrical coordinate system $(R,z,\psi)$ is 
used with
the azimuthal angular coordinate $\psi$ counted from the $l=0^o$ plane;
the major axis of the perturbation is along $\psi=\alpha$ with
amplitudes ${B_m(R,z) \over \sin m\alpha}$, where $B_m(R,z)$ specifies
the odd part of the density (cf. eq.~\ref{oddi2}), and $C(R,z)$ is an 
axisymmetric fit to the addition map (cf. eq.~\ref{eveni2}), 
and $K_m(R,z,\psi)$ is an invisible density, made by subtracting 
an axisymmetric component $V_m(R,z)$ from an end-on bar 
$B_m(R,z) \cos m\psi$.

A few brief comments before proceeding to derive the limits on
the bar angle.   Eqs.~\ref{num}-~\ref{kmin} imply that any triaxial bar 
with the same  amplitudes $C(R,z)$ and $B_m(R,z)$
will follow a sequence with
identical line-of-sight integrated light distributions on both positive 
and negative $l$ sides $I(\pm l,b)$,
the same odd parts $\nu_{odd}$
(with respect to the $\psi=l=0^o$ plane) in density,
and different even parts $\nu_{even}$, but with the difference being
a linear superposition of invisible densities
$K_m(R,z,\psi)$.  For triaxial densities with generally
multiple modes ($m=2,4,...$), they are related to each other by an
invisible density 
\beq\label{kchain}
\sum_{m=2,4,...} (\cot m\alpha'-\cot m\alpha)K_m(R,z,\psi),
\eeq
where {\sl the sum of different modes in the given proportion is
necessary to prevent spiral-arm-like twist of the major axis}.
Adding this invisible density
to a model with symmetry plane at $\psi=\alpha$ rotates the plane to
$\psi=\alpha'$.
The sequence of edge-on models given by eq.~\ref{nuf1} and ~\ref{edgeonf}
is a subclass of the models here with $m=2$, and
\beq\label{bvrz}
B_2(R,z) = -c_1 P(r) {R^2 \over r^2},~
V_2(R,z) = -2 c_1  S(r) - B_2(R,z),
\eeq
\beq\label{crz}
C(R,z) = G(r) + (c_0-c_2) S(r) + P(r) \left(s_0-c_2 + 2 c_2  {R^2 \over r^2}\right).
\eeq
Generally $C(R,z)$ and $B_m(R,z)$ should be derived from fitting the 
addition and the difference maps of the projected light,
\bey\label{eveni2}
{1 \over 2} \left[I(l, b) + I(-l, b)\right] 
&=&\int_0^\infty  \! \! C(R,z)  dD,\\\label{oddi2}
{1 \over 2} \left[I(l, b) - I(-l, b)\right] 
&=&\int_0^\infty  \! \! B_m(R,z) \sin m \psi dD,
\eey
and $V_m(R,z)$ from eq.~\ref{km}.

Impose positivity requirements to $\nu_{\alpha}(R,z,\psi)$,
\beq
A_{\alpha}(R,z) \equiv \left[ C(R,z) -\cot m\alpha V_m(R,z) \right]  \ge 
{B_m(R,z) \over \sin m\alpha}
\eeq
for all $(R,z)$.  This reduces to an upper and a lower bound for 
the angle $\alpha$, 
\beq\label{arange}
\Max\left[t_{-}\right] \le \tan{m \alpha \over 2} \le \Min\left[t_{+}\right],
\eeq
where $t_{-} \le t \le t_{+}$ is the range
bounded by the effectively quadratic inequality for a variable $t$
\beq\label{tpm}
{B_m - V_m \over 2 C} t +{B_m + V_m \over 2 C} t^{-1} \le 1
\eeq
at a given position $(R,z)$ on the meridional plane,
and the overlapped interval of these ranges is used in eq.~\ref{arange}.  
The exact upper and lower bounds of the angle $\alpha$ 
for the COBE bar 
will be discussed elsewhere
as it involves solving three integral equations numerically
(eqs.~\ref{eveni2} and~\ref{oddi2} for $B_m(R,z)$ and $C_m(R,z)$
and eq.~\ref{km} for $V_m(R,z)$), and several practical issues 
not considered so far 
(including a detailed dust model for the COBE/DIRBE maps, 
the choice of fitting regions with reliable photometry, the degree of
smoothness of the solution in a non-parametric fit, and 
search of solutions in the 1D parameter space spanned by $\alpha$).
Here it is only worth commenting on two points.  First
Binney, Gerhard \& Spergel (1997) showed that it is possible to obtain good
fits to the COBE maps with a triaxial bar if the bar angle is in
the range $15^o \le \alpha \le 35^o$ and the Sun is
very close the midplane of the bar.  A sequence of models which is
roughly consistent with these limits is discussed in the next section.
Second the range for $\alpha$ is set by $C_m$ and $B_m$,
which are in turn set by the addition map and the difference map
respectively with an integration over the line of sight 
(cf. eq.~\ref{eveni2} and~\ref{oddi2}).  Hence
the range for $\alpha$ is also a function of the distance to the bar.
However, the upper and lower limits are only weakly dependent on the
distance as long as the observer is outside the bar, as shown 
in Fig.3 by the nearly horizontal boundaries of the shaded region 
for $R_0 \gg 2$ kpc.

\begin{figure}
\epsfysize=5cm
\leftline{\epsfbox{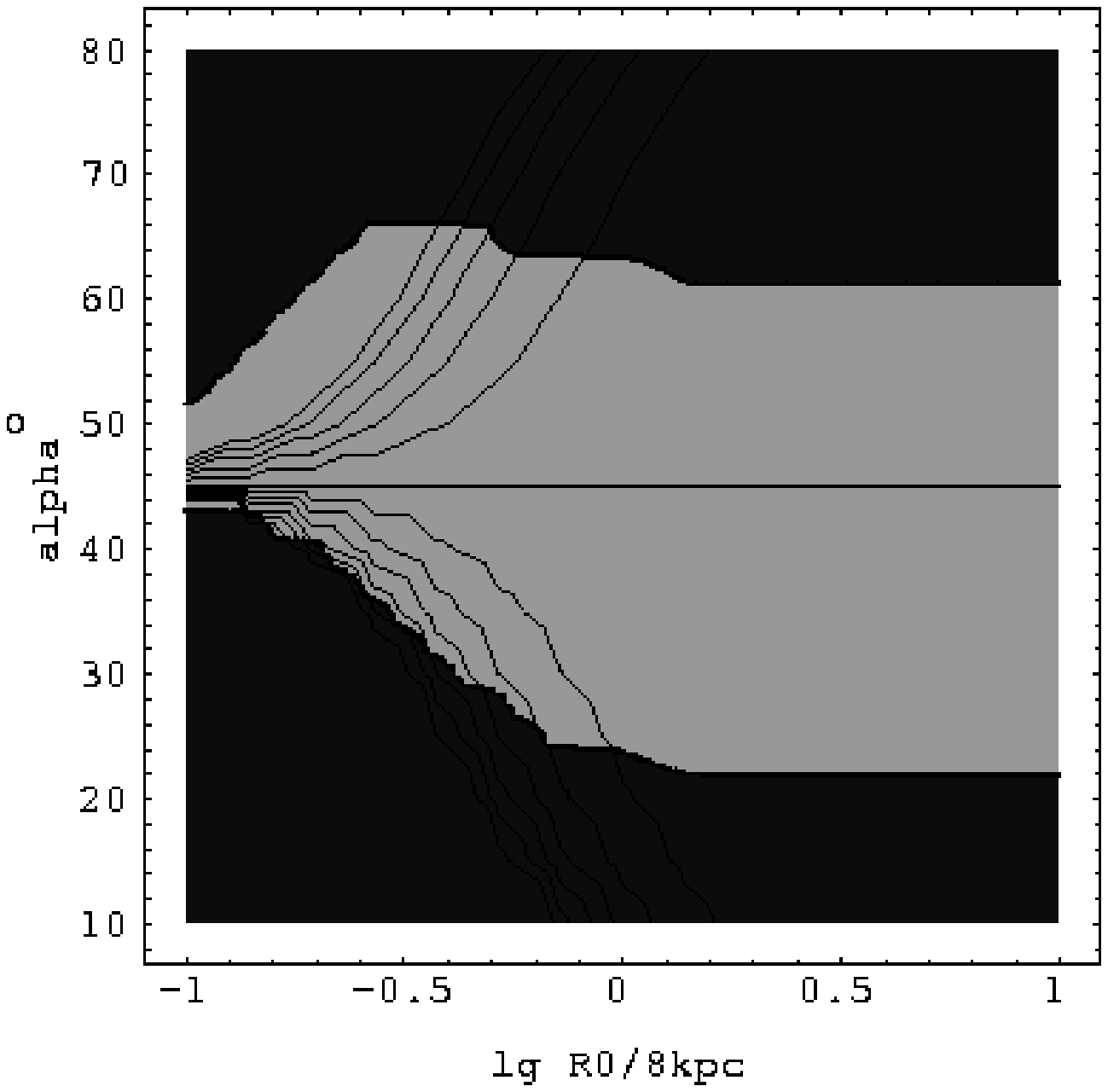}}
\vskip 0.5cm
\epsfysize=5cm
\leftline{\epsfbox{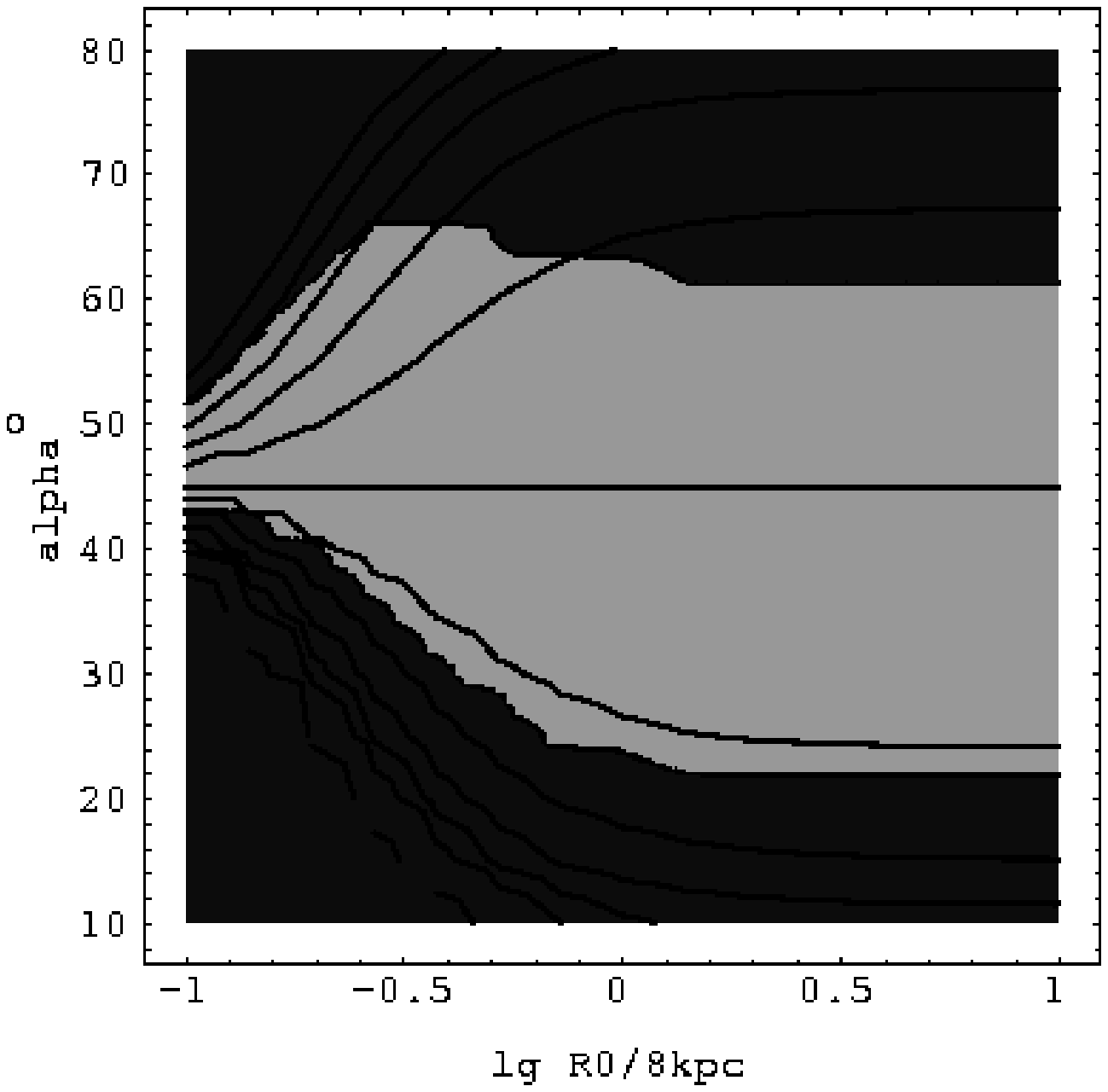}}
\caption{
For an edge-on bar with a fixed $M/L$ and parameters given in 
eq.~\ref{examp}, positivity of the density puts
loose upper and lower limits on the bar 
angle (cf. eqs.~\ref{arange},~\ref{tpm},~\ref{bvrz} and~\ref{crz}).
These limits are functions of the distance
to the bar, which mark the upper and lower boundaries of the shaded region
in the $\log (R_0/8{\rm kpc})$ vs. $\alpha$ plane.  
The uncertainty of the bar angle propagates to 
the inferred value for the luminosity of the bar.
The contours show the bar's absolute magnitude (upper panel)
in the $\log (R_0/8{\rm kpc})$ vs. $\alpha$ parameter plane
with an interval of 0.05 magnitude, and
the equal contours of 
the central escape velocity $\sqrt{2|\Phi(0)|}$ (lower panel)
in intervals of 5\%.  Both the escape velocity and the luminosity
increase with increasing $\alpha$, but the increment is small
as long as one is observing from well outside
the bar, despite the strong ambiguity with 
the angle of the bar.}
\end{figure}

\subsection{A toy model and implications to the COBE bar}

To illustrate the non-uniqueness in the Galactic bar, and how to lift
the degeneracy, consider edge-on models in eq.~\ref{nuf} and~\ref{Hf} with
\beq\label{examp}
a=2~{\rm kpc},~c_0=c_2=-{c_1 \over 2}={1 \over 6},~s_0=-1,
\eeq
and (cf. eq.~\ref{edgeonf})
\beq
s_1=s_2=0,~~f=c_2-c_0 - 2 c_1 \cot 2 \alpha = {2 \over 3}\cot 2 \alpha,
\eeq 
which prescribes a sequence of edge-on $m=2$ bars 
with a Gaussian radial profile inside about 4 kpc of the center 
with major axis at an angle $\alpha$ from the Sun-center line.  
The density is strictly positive
everywhere for models with a broad range of major axis angles,
$22.5^o \le \alpha \le 63.5^o$ 
for $R_0=8$ kpc (cf. the shaded region in Fig. 3).
Models are also monotonic everywhere except for a small bump
near ${R_0\over 2} \sim 4$ kpc on the short and middle axes.
These limits will be significantly relaxed when an axisymmetric exponential 
disc is included because it adds a positive background distribution;
a disc with peak intensity about $1/10$ of the bar can also make 
$\alpha \sim 15^o$ and $\alpha \sim 75^o$ models positive everywhere.

While the specific models here are clearly toy models, too simplistic
to fit the COBE map in any detail, as shown in Fig.2, they grossly
resemble the Galactic bar in terms of its Gaussian radial profile
(Dwek et al. 1995), its shape and axis ratios.  In particular, the
range of the bar angle here is largely in agreement with those in the
literature (see a table compiled in Gerhard 1995), including the
results of Binney, Gerhard \& Spergel (1997) and Dwek et al. (1995)
derived from fitting the COBE map, and Stanek et al. (1994, 1997)
derived from fitting the star count data of bulge red clump giants.
Hence the models are useful to demonstrate the non-uniqueness and test
various techniques for lifting the non-uniqueness.

Fig.2 shows an equatorial slice of the model at $\alpha=25^o$ (the
thin peanut-shaped bar) in heavy dashed contours and a model at
$\alpha=50^o$ (the more concentrated bar) in solid contours.
These two grossly different models look exactly the same in
line-of-sight integrated intensity for an observer at $(-1,0)$ at the
central left (the assumed position of the Sun).  Neither of these two
models can be easily ruled out on the basis of stability and
self-consistency.  

Nevertheless it is possible to break the
degeneracy with kinematic data since {\sl the invisible density
changes the orientation of the bar potential model as well as its
shape and depth.}  For the models in eq.~\ref{nuf}, the
depth of the potential well at the center is
\beq
\left|\Phi(0)\right| = \pi \nu_0 a^2 G(M/L) 
\left[\left( 4 + 2s_0 + {2 \over 3}c_0 \right)
- \left( {1 \over 3} + {a^2 \over R_0^2} \right) f \right],
\eeq
which decreases linearly with increasing fraction ($f$) of the superimposed 
invisible density, where $M/L$ is the mass-to-light ratio, assumed to be
the same for all models.
As a result, the potential well of the $\alpha=25^o$ model 
is slightly shallower 
than that of the more centrally concentrated $\alpha=50^o$ model (cf Fig.2).
However, the difference is only 7\% in term of the maximum escape velocity
$\sqrt{2|\Phi(0)|}$ of the models (cf. lower panel of Fig.3).
The differences in terms of
the mass weighted average velocity dispersion (estimated from the virial
theorem) and the circular velocity (estimated directly from the potential) 
are also at only a few percent level, too small to be measured with certainty.
To distinguish the models with kinematic data,  one needs to model the
anisotropy and perspective effects of the bar orbits in detail, e.g., 
the longitude-velocity diagram for orbits of gas clouds
(Binney et al. 1991, Zhao, Rich \& Spergel 1996), 
and the vertex deviation in the stellar velocity
ellipsoid (Zhao, Spergel \& Rich 1994); these models introduce
additional complexity such as the pattern speed of the bar, and
the stellar distribution function.

The most straightforward approach to break the degeneracy of the COBE
bars is perhaps to compare the models with measurements of the line of
sight distance distribution of the bar from color-magnitude diagrams.
The depth of the models, defined here as the rms dispersion of
distance in the Sun-center line of sight, decreases from $0.26R_0$ for
$\alpha=22.5^o$ to $0.19R_0$ for $\alpha=63.5^o$.  The 30\% difference
is detectable with good distance indicators and large data sets, such
as the bulge red clump giants in the microlensing and variable star
surveys towards the Galactic bulge (Paczy\'nski 1996).  It is also
interesting that {\sl the invisible density also leaves observable
traces on the microlensing maps}, i.e., the distribution of the
optical depth on the sky plane (Evans 1995, Bissantz, Englmaier,
Binney \& Gerhard 1997).  This is because the optical depth is
proportional to both the projected intensity and the line-of-sight
depth, and while an invisible density does not change the former, it
changes the latter.  The microlensing optical depth increases by about
$50\%$ by going from the $\alpha=63.5^o$ model to the $\alpha=22.5^o$
model.  Nevertheless, since the data set on bulge red clump giants is
on the order of a million times larger than that of microlensed
sources, it argues for the former plus the COBE map being the speedest
approach.  However, fitting the microlensing optical depth,
gas/stellar velocities can set the normalization of the density,
namely the mass-to-light ratio.  To fit the same projected light
intensity the models should have the same normalization
$\nu_0$\footnote{ $\nu_0$ is not the peak density $\nu(0)= \left(1-{f
\over 2}\right)\nu_0$ (cf. eq.~\ref{nuf}), which roughly doubles by
going from a $\alpha=22.5^o$ model to a $\alpha=63.5^o$ model.}, but
the total luminosity increases by about $8\%$ (cf.eq. ~\ref{lum}) by
going from a $\alpha=22.5^o$ model to a $\alpha=63.5^o$ model.

\begin{figure}
\epsfysize=7cm
\epsfbox{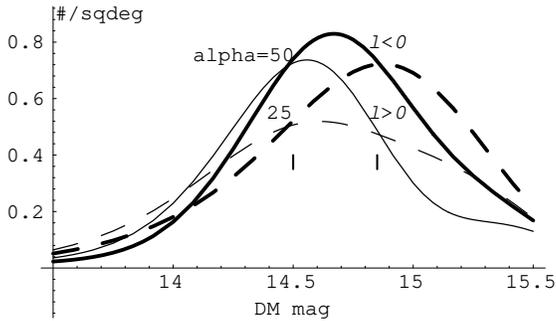}
\caption{
Simulated distance modulus distributions for red clump giants 
in the two bar models $\alpha=25^o$ (dashed curves) and $50^o$ (solid curves) 
in Fig. 2 at
line of sight directions $(\pm 5^o, -4^o)$. }
\end{figure}

Fig.4 shows the simulated star count data for the two models at line
of sight directions $(l,b)=(\pm 5^o,-4^o)$, with an assumed
Galactocentric distance modulus (DM) $14.5$ magnitude.  The
predictions have taken into account the growing volume with radius for
a given solid angle, and have been convolved with the dispersion of
the absolute magnitude of the clump giants (about $0.2$ magnitude).
The predicted distance modulus distributions are quite different for
the two models in terms of the peak position, the width and shape of
profiles.  The fact that for both models the $l>0$ side is closer than
the $l<0^o$ side is because the major axes of both bars are placed at
positive $\alpha$.  Interestingly the amount of $l>0^o$ vs. $l<0^o$
asymmetry is the same for both models, as indicated by the two short vertical
lines in Fig.4, because the left-to-right symmetric invisible
density has no effect on the asymmetry.  The distributions resemble
Fig. 2 of Stanek et al. (1994), which show the apparent magnitude
distribution of Galactic bulge red clump giants at the same
low-extinction line of sight directions $(\pm 5^o, -4^o)$ in the OGLE
microlensing surveys.  While it is premature to use the simple
analytical models here to draw any conclusion on the bar angle, they
suggest that an integration of the COBE map with the star count data
of bulge red clump giants (Stanek et al. 1997) can lift the degeneracy
of the bar models.

\section{Conclusion}

The non-uniqueness in deprojecting the COBE/DIRBE maps originates from
the simple fact that a round bulge can mimic an end-on bar in the
line-of-sight integrated light distribution.  This kind of
non-uniqueness is almost lifted when bar-like reflection symmetries
are imposed, except that the angle of these symmetry planes with our
line of sight is still a free parameter, which can only be loosely
constrained by the positivity of the deprojected density.  The main
results are summarized as follows.  (1) Triaxial Galactic bar models,
which project to identical intensity map in the absence of extinction,
form a 1D sequence as function of the bar's orientation, in agreement
with numerical results of Binney \& Gerhard (1996).  (2) It is
ill-conditioned to speak of any best-fitting bar model because the
limit on the bar parameters comes from plausibility of the density
(positivity, smoothness, and regular-looking shape and radial profile,
etc.), not from the goodness of fit.  (3) The theory predicts that
COBE models likely follow a sequence with identical $\nu_{odd}$, which
is uniquely constrained by the perspective effects of a triaxial bar.
(4) The non-uniqueness in the even part of the density $\nu_{even}$ is
of nature entirely different from those induced by shearing/stretching
of an external ellipsoid (Stark et al. 1977) or adding a konus-like
structure (Gerhard \& Binney 1996).  (5) Various methods to fully
break the degeneracy of COBE bars are compared.  In agreement with
findings for extragalactic axisymmetrical systems (van den Bosch 1997,
Romanowsky \& Kochanek 1997), it is unlikely to put a very tight limit
on the bar's parameter space with general dynamical constraints such
as self-consistency and stability, but it is still hopeful to lift
some degeneracy by fitting kinematic data of gas and stars with
sophisticated dynamical models.  With the increasing number of
microlensing events towards the bulge, microlensing optical depth maps
also have the potential of lifting the degeneracy of the models.  But the
cleanest approach to remove the ambiguity of the bar angle is to
compare with star count data from the current bulge microlensing and
variable star surveys, in particular the large data set on the bulge
red clump giants, which provide the extra constraints on the ``depth''
of the bar (Stanek et al. 1997).  Extinction is the main limiting
factor in analysis of both star counts and COBE maps.  Finally the
mass-to-light ratio, the last parameter of a complete mass model of
the bar, can be determined from fitting gas/stellar kinematics or the
microlensing optical depth (Bissantz, Englmaier, Binney, \& Gerhard
1997).

I thank Frank van den Bosch for a helpful discussion in the early
stage of this work, Tim de Zeeuw and Ortwin Gerhard for a careful
reading of an early draft.

{}

\bsp
\label{lastpage}
\end{document}